\documentclass[a4paper]{report}
\usepackage[utf8]{inputenc}
\usepackage[T1]{fontenc}
\usepackage{RJournal}
\usepackage{amsmath,amssymb,array}
\usepackage{booktabs}

\providecommand{\tightlist}{%
  \setlength{\itemsep}{0pt}\setlength{\parskip}{0pt}}

\usepackage{amsmath, amsthm, amsfonts, amssymb, bm, natbib}

\begin{document}

\begin{article}
\title{The openVA Toolkit for Verbal Autopsies}
\author{
\begin{centering}
Zehang Richard Li$^1$, Jason Thomas$^2$, Eungang Choi$^2$, Tyler H.
McCormick$^{3}$, and Samuel J Clark$^2$
\\[1em]
\small 1. Department of Statistics, University of California, Santa Cruz.\\
\small 2. Institute for Population Research and Department of Sociology, Ohio State University \\
\small 3. Department of Statistics and Department of Sociology, University of Washington\\
\end{centering}
}

\maketitle

\abstract{%
Verbal autopsy (VA) is a survey-based tool widely used to infer cause of
death (COD) in regions without complete-coverage civil registration and
vital statistics systems. In such settings, many deaths happen outside
of medical facilities and are not officially documented by a medical
professional. VA surveys, consisting of signs and symptoms reported by a
person close to the decedent, are used to infer the COD for an
individual, and to estimate and monitor the COD distribution in the
population. Several classification algorithms have been developed and
widely used to assign causes of death using VA data. However, the
incompatibility between different idiosyncratic model implementations
and required data structure makes it difficult to systematically apply
and compare different methods. The \CRANpkg{openVA} package provides the
first standardized framework for analyzing VA data that is compatible
with all openly available methods and data structure. It provides an
open-source, R implementation of several most widely used VA methods. It
supports different data input and output formats, and customizable
information about the associations between causes and symptoms. The
paper discusses the relevant algorithms, their implementations in R
packages under the \CRANpkg{openVA} suite, and demonstrates the pipeline
of model fitting, summary, comparison, and visualization in the R
environment.
}

\hypertarget{introduction}{%
\section{Introduction}\label{introduction}}

Verbal autopsy (VA) is a well-established approach to ascertaining the
cause of death (COD) when medical certification and full autopsies are
not feasible or practical
\citep{garenne2014prospects-BMC, taylor1983child}. After a death is
identified, a specially-trained fieldworker interviews the caregivers
(usually family members) of the decedent. A typical VA interview
includes a set of structured questions with categorical or quantitative
responses and a narrative section that records the ``story'' of the
death from the respondent's point of view \citep{world2012verbal}.
Currently, there are multiple commonly used questionnaires with
overlapping, but not identical questions. VAs are routinely used both by
researchers in health and demographic surveillance systems
\citep{Sankoh2012, maher2010translating} and multi-country research
projects \citep{nkengasong2020improving, breiman2021postmortem}, and in
national-scale surveys in many low- and middle-income countries (LMICs).
For a more comprehensive overview of the current use of VA, we refer
readers to \citet{Chandramohan2021Estimating}.

The process of inferring a cause from VA data consists of two
components. First, there must be some external information about the
relationship between causes and symptoms. In supervised learning
problems, the external information is typically derived from training
datasets with known labels. In the context of VA, a common practice is
to obtain training data using clinically trained, experienced physicians
who read a fraction of the interviews and determine causes. To address
the fact that physicians frequently do not agree on causes, VA
interviews are often read by two physicians, and sometimes three, and
the final causes are determined through a consensus mechanism
\citep[e.g.,][]{kahn2012profile}. This process can be extremely time-
and resource-intensive. Another means of obtaining this information is
directly through expert opinion. For example, one can ask groups of
physicians to rate the likelihood of a symptoms occurring, given a
particular COD, which can be converted into a set of probabilities of
observing a symptom given a particular cause. Such expert knowledge can
be highly useful in analyzing VA data when training datasets do not
exist.

Secondly, an algorithmic or statistical model is needed to assign the
causes of death by extrapolating the relationship between symptoms and
causes to the target population. The cause-of-death assignment model is
conceptually a separate construct from the symptom-cause information.
The current state of the VA literature, however, often does not
distinguish between the two. This is in part due to popular VA software
that combines a source of symptom-cause information (e.g., a certain
training data or a database of expert knowledge) with a specific
algorithm, and requires a specific type of survey instrument to be used.
This restriction prevents robust comparison between methods and
contexts. A health agency in one region may, for example, want to
analyze VA data using the same VA algorithm as a neighboring region to
ensure estimates are comparable. Unless the agencies used the same
survey format, however, this is not possible with existing tools.

The \CRANpkg{openVA} package \citep{openvapkg} addresses this issue
through an open-source toolkit. The \CRANpkg{openVA} suite comprises a
collection of R packages for the analysis of verbal autopsy data. The
goal of this package is to provide researchers with an open-source tool
that supports flexible data input formats and allows easier data
manipulation and model fitting. The \CRANpkg{openVA} suite consists of
four core packages that are on CRAN, \CRANpkg{InterVA4}
\citep{interVA4}, \CRANpkg{InterVA5} \citep{interVA5},
\CRANpkg{InSilicoVA} \citep{insilicovapkg}, and \CRANpkg{Tariff}
\citep{tariffpkg}, and an optional package \pkg{nbc4va} \citep{nbcpkg}.
Each of these packages implements one coding algorithm. For three of
these algorithms -- namely, InterVA-4, InterVA-5 and Tariff -- there are
also compiled software programs distributed by the original authors.

The main focus of this paper is to provide a general introduction to the
implementation details of the included algorithms both in terms of the
underlying methodology, and through case studies. The \CRANpkg{openVA}
package has four major contributions:

\begin{enumerate}
\def\labelenumi{\arabic{enumi}.}
\tightlist
\item
  It provides a standardized and user-friendly interface for analysts to
  fit and evaluate each method on different types of data input using
  standard syntax. Previously, most of the methods were designed to be
  used specifically with their own input formats and are usually
  incompatible with others. The \CRANpkg{openVA} package closes this gap
  by allowing easier and fair model comparison of multiple algorithms on
  the same data. This significantly facilitates further research on VA
  algorithms.
\item
  It provides a series of functionalities to summarize and visualize
  results from multiple algorithms, which is helpful for analysts not
  familiar with data manipulation and visualization in R.
\item
  It does not directly implement any algorithms for coding VA data, so
  that it is possible for a research group to maintain their own
  algorithm implementations callable from the \CRANpkg{openVA} package,
  while also making it available to general users as a standalone piece
  of software. For example, the \pkg{nbc4va} was developed and
  maintained independently by researchers at the Center for Global
  Health Research in Toronto, but is designed so that it can be
  seamlessly integrated into the \CRANpkg{openVA} package.
\item
  It is fully open source, and can be run on multiple platforms. The
  open-source nature of \CRANpkg{openVA} significantly expands its
  potential for methodological research and its suitability for
  integration within a larger data analysis pipeline. Compared to the
  alternative implementations, the InterVA-4 and InterVA-5 software are
  distributed as compiled software that can only be run on Windows
  system. They provide the source codes as an additional code script,
  which are difficult to modify and re-compile. Tariff, as implemented
  through the SmartVA-Analyze application \citep{serina2015improving},
  was also primarily distributed as a compiled application that can only
  be run on Windows system \citep{smartVA-web}. However, their source
  codes were recently made available under the open source MIT License
  on GitHub \citep{smartVA-git}.
\end{enumerate}

The rest of this paper is organized as follows: We first briefly
introduce the main component packages and the underlying algorithms. We
then demonstrate standard data structures and model fitting steps with
the \CRANpkg{openVA} package and functionalities to summarize results.
We then discuss how additional information can be incorporated into
modeling VA data, and we briefly survey additional packages and software
developments built around the \CRANpkg{openVA} package. We end with a
discussion of remaining issues and limitations of the existing automated
VA algorithms and propose new functionalities to be included in
\CRANpkg{openVA} package.

\hypertarget{structure-of-the-openva-package}{%
\section{Structure of the openVA
package}\label{structure-of-the-openva-package}}

The \CRANpkg{openVA} suite of packages currently consists of four
standalone packages that are available on CRAN and one optional package
hosted on GitHub. In this section, we first provide a brief introduction
to these five packages, and we discuss the mathematical details behind
each algorithm in the next subsection.

\begin{itemize}
\item
  \CRANpkg{InterVA4} \citep{li2014interva4, interVA4} is an R package
  that implements the InterVA-4 model \citep{byass2012strengthening}. It
  provides replication of the widely used InterVA software
  \citep{2015interVA}. The standard input of \CRANpkg{InterVA4} is in
  the form of a pre-defined set of indicators, based on the 2012 World
  Health Organization (WHO) VA instrument \citep{world2012verbal}. The
  default InterVA-4 algorithm cannot be applied to other data input
  formats because its internal built-in prior information is specific to
  a fixed set of indicators and causes. The same restriction is also
  maintained in the \CRANpkg{InterVA4} package. However, the
  mathematical formulation of the InterVA-4 model is completely
  generalizable to other binary input formats, as described in later
  sections, and is also implemented in the \CRANpkg{openVA} package.
\item
  \CRANpkg{InterVA5} \citep{interVA5} is an R package that implements
  the InterVA-5 model \citep{byass2019integrated}. The InterVA-5 model
  updates the previous version in several ways. First, the input data
  must adhere to the format of the 2016 WHO VA instrument
  \citep{dambruoso2017}. Second, changes have been made to the data
  processing steps. It is also worth noting that the model outputs have
  been expanded by the inclusion of the most likely Circumstances Of
  Mortality CATegory, or COMCAT, among the results -- the categories
  include: culture, emergency, health systems, inevitable, knowledge,
  resources, or an indeterminant combination of multiple factors
  \citep[for more details, see][]{dambruoso2017}. Despite these changes,
  the mathematical formulation of InterVA-5 is identical to that of
  InterVA-4.
\item
  \CRANpkg{InSilicoVA} \citep{insilicovapkg} is an R package that
  implements the InSilicoVA algorithm, a Bayesian hierarchical framework
  for cause-of-death assignment and cause-specific mortality fraction
  estimation proposed in \citet{insilico}. It is originally designed to
  work with the WHO VA instrument, i.e., the same input data used by the
  InterVA software, but is also generalizable to other data input
  formats. It is a fully probabilistic algorithm and can incorporate
  multiple sources of information, such as known sub-populations in the
  dataset, and physician coding when such data are available. The Markov
  Chain Monte Carlo (MCMC) sampler is implemented in Java for improved
  speed.
\item
  \CRANpkg{Tariff} \citep{tariffpkg} is an R package that implements the
  Tariff algorithm \citep{james}. It most closely reflects the
  description of the Tariff 2.0 method \citep{serina2015improving}. The
  Tariff algorithm is developed by the Institute for Health Metrics and
  Evaluation (IHME) and is officially implemented in the SmartVA-Analyze
  software \citep{smartVA-web}. However, as the developers of this R
  package are not affiliated with the authors of the original algorithm,
  there are some discrepancies in implementation. The source code of the
  two versions of Tariff was not publicly available at the time when the
  \CRANpkg{Tariff} package was created, so the package was developed
  based solely on the descriptions in the published work. Despite the
  difference in implementation, \CRANpkg{Tariff} is able to achieve
  comparable results as the published work as demonstrated in
  \citet{insilico}. More detailed descriptions of the \CRANpkg{Tariff}
  implementations are also discussed in the supplement of
  \citet{insilico}. The later released Python source codes of
  SmartVA-Analyze have been incorporated in the web application
  extension of the \CRANpkg{openVA} package.
\item
  \pkg{nbc4va} \citep{nbcpkg} is an R package that implements the Naive
  Bayes Classifier for VA encoding \citep{miasnikof2015naive}. It
  calculates the conditional probabilities of symptoms given causes of
  death from a training dataset, instead of using physician-provided
  values. \pkg{nbc4va} is developed and maintained by researchers at the
  Center for Global Health Research in Toronto, but is designed so that
  it can be seamlessly integrated into \CRANpkg{openVA}. Currently, the
  \pkg{nbc4va} package is hosted on GitHub and is an optional package
  that users can choose to load separately.
\end{itemize}

We note that there are additional methods and implementations to assign
causes of death using VA data that are not included in the
\CRANpkg{openVA} implementation
\citep[e.g.,][]{flaxman2011random, jeblee2019automatically}. These
methods are not widely adopted by VA practitioners and most of these
implementations are either not publicly available or require data
processing steps that are specific to certain datasets, making them
impractical for routine use in general. In addition, while the
cause-of-death assignment process is closely related to the generic
multi-class classification problem, naive application of off-the-shelf
classification algorithms has been shown to perform poorly in the
context of VA \citep{murray2014using}. Therefore, we focus on the
methods that are currently adopted by practitioners. We briefly survey
some more recent developments and their implementations at the end of
this paper.

The \CRANpkg{openVA} package is hosted on CRAN and can be installed with
the following commands. Since posterior inference is carried out using
MCMC in the InSilicoVA algorithm, we set the seed for the random number
generator to make the paper reproducible. For the analysis in this
paper, we also install the \pkg{nbc4va} package separately from GitHub.
The versions of the supporting packages can be checked in R using the
\texttt{openVA\_status()} function.

\begin{Schunk}
\begin{Sinput}
set.seed(12345)
library(openVA)
remotes::install_github("rrwen/nbc4va")
library(nbc4va)
openVA_status()
\end{Sinput}
\end{Schunk}

\hypertarget{overview-of-va-cause-of-death-assignment-methods}{%
\subsection{Overview of VA cause-of-death assignment
methods}\label{overview-of-va-cause-of-death-assignment-methods}}

The common modeling framework for VA data consists of first converting
the survey responses into a series of binary variables, and then
assigning a COD to each death based on the binary input variables.
Typically, the target of inference consists of two parts: the individual
cause-of-death assignment, and the population-level cause-specific
mortality fractions (CSMF), i.e., the fraction of deaths due to each
cause. In this section, we formally compare the modeling approaches
utilized by each algorithm for these two targets. We adopt the following
notations. Consider \(N\) deaths, each with \(S\) binary indicators of
symptoms. Let \(s_{ij}\) denote the indicator for the presence of
\(j\)-th symptom in the \(i\)-th death, which can take values 0, 1, or
NA (for missing data). We consider a pre-defined set of causes of size
\(C\). For the \(i\)-th death, denote the COD by
\(y_i \in \{1, ..., C\}\) and the probability of dying from cause \(k\)
is denoted by \(P_{ik}\). For the population, the CSMF of cause \(k\) is
denoted as \(\pi_k\), with \(\sum_{k=1}^C \pi_k = 1\).

\begin{itemize}
\item
  \textbf{InterVA4} \citep{byass2012strengthening} and \textbf{InterVA5}
  \citep{byass2019integrated} algorithms calculate the probability of
  each COD given the observed symptoms using the following formula, \[
  P_{ik} = \frac{\pi_{k}^{(0)} \prod_{j=1}^S P(s_{ij}=1|y_{i}=k) \mathbf{1}_{s_{ij} = 1}}
  {\sum_{k' = 1}^C \pi_{k'}^{(0)} \prod_{j=1}^S P(s_{ij}=1|y_{i}=k') \mathbf{1}_{s_{ij} = 1}}
  \] where both the prior distribution of each of the causes,
  \(\pi_{k}^{(0)}\) and the conditional probabilities
  \(P(s_{ij} = 1 | y_i = k)\) are fixed values provided in the InterVA
  software. It is worth noting that the formula does not follow the
  standard Bayes' rule as it omits the probability that any symptom is
  absent. A detailed discussion of this modeling choice can be found in
  \citet{insilico}. The conditional probabilities,
  \(P(s_{ij}=1|y_{i}=k)\), used in InterVA algorithms are represented as
  rankings with letter grades instead of numerical values
  \citep{byass2012strengthening}. For example,
  \(P(s_{ij}=1|y_{i}=k) = A+\) is translated into
  \(P(s_{ij}=1|y_{i}=k) = 0.8\), etc. The standard InterVA software only
  supports the fixed set of symptoms and causes where such prior
  information is provided. For a different data input format, this
  formulation can be easily generalized if training data are available.
  We include in the \CRANpkg{openVA} package an extension of the
  algorithm that calculates \(\hat P(s_{ij}=1|y_{i}=k)\) from the
  empirical distribution in the training data and then maps to letter
  grades with different truncation rules. Details of the extended
  InterVA algorithm can be found in \citet{insilico}.

  After the individual COD distributions are calculated, InterVA-4
  utilizes a series of pre-defined rules to identify up to the top three
  most likely COD assignments, and truncates the probabilities for the
  rest of the CODs to 0 and adds an `undetermined' category so that the
  probabilities sum up to 1 (See the user guide of \citet{2015interVA}).
  Then the population-level CSMFs are calculated as the aggregation of
  individual COD distributions, such that \[
  \pi_k = \sum_{i=1}^N P^*_{ik}
  \] where \(P^*_{ik}\) denotes the individual COD distribution after
  introducing the undetermined category.
\item
  \textbf{Naive Bayes Classifier} \citep{miasnikof2015naive} is very
  similar to the InterVA algorithm with two major differences. First,
  instead of considering only symptoms that present, the NBC algorithm
  also considers symptoms that are absent. Second, the conditional
  probabilities of symptoms given causes are calculated from training
  data instead of given by physicians, which is similar to our extension
  of InterVA discussed above. Similar to InterVA, the NBC method can be
  written as \[
  P_{ik} = \frac{\pi_{k}^{(0)} \prod_{j=1}^S (P(s_{ij}=1|y_{i}=k) \mathbf{1}_{s_{ij} = 1} + P(s_{ij} \neq 1|y_{i}=k) \mathbf{1}_{s_{ij} \neq 1})}
  {\sum_{k' = 1}^C \pi_{k'}^{(0)} \prod_{j=1}^S (P(s_{ij}=1|y_{i}=k') \mathbf{1}_{s_{ij} = 1}+ P(s_{ij} \neq 1|y_{i}=k') \mathbf{1}_{s_{ij} \neq 1})}
  \] and the CSMFs are calculated by \(\pi_k = \sum_{i=1}^N P_{ik}\).
\item
  \textbf{InSilicoVA} algorithm \citep{insilico} assumes a generative
  model that characterizes both the CSMF at the population level, and
  the COD distributions at the individual level. In short, the core
  generative process assumes \begin{eqnarray} 
  s_{ij} | y_i = k &\propto& \mbox{Bernoulli}(P(s_{ij} | y_i = k)) \\
  y_i | \pi_1, ..., \pi_C &\propto& \mbox{Categorical}(\pi_1, ..., \pi_C) \\
  \pi_k &=& \exp \theta_k / \sum_{k=1}^C \exp \theta_k \\
  \theta_k &\propto& \mbox{Normal}(\mu, \sigma^2)
  \end{eqnarray} and hyperpriors are also placed on
  \(P(s_{ij} | y_i = k)\), \(\mu\), and \(\sigma^2\). The priors for
  \(P(s_{ij} | y_i = k)\) are set by the rankings used in InterVA-4 if
  the data are prepared into InterVA format, otherwise they are learned
  from training data. The priors on \(\mu\) and \(\sigma^2\) are diffuse
  uniform priors. Parameter estimation is performed using MCMC, so that
  a sample of posterior distributions of \(\pi_k\) can be obtained after
  the sampler converges.
\item
  \textbf{Tariff} algorithm \citep{james} differs from the other three
  methods in that it does not calculate an explicit probability
  distribution of the COD for each death. Instead, for each death \(i\),
  a Tariff score is calculated for each COD \(k\) so that \[
  Score_{ik} = \sum_{j = 1}^{S} \mbox{Tariff}_{kj}\mathbf{1}_{s_{ij}=1}
  \] where the symptom-specific Tariff score \(\mbox{Tariff}_{kj}\) is
  defined as \[
  \mbox{Tariff}_{kj} = \frac{n_{kj} - median(n_{1j}, n_{2j}, ..., n_{Cj})} {IQR(n_{1j}, n_{2j}, ..., n_{Cj})}
  \] where \(n_{kj}\) is the count of how many deaths from cause \(k\)
  contain symptom \(j\) in the training data. The Tariff scores are then
  turned into rankings by comparing them to a reference distribution of
  scores calculated from re-sampling the training dataset to obtain a
  uniform COD distribution. It is worth noting that the Tariff algorithm
  produces the COD distribution for each death in terms of their
  rankings instead of the probability distributions. And thus the CSMF
  for each cause \(k\) is calculated by the fraction of deaths with
  cause \(k\) being the highest ranked cause, i.e., \[
  \pi_k = \frac{\sum_{i=1}^N\mathbf{1}_{y_i = k}}{N}
  \]
\end{itemize}

In addition to the different model specifications underlying each
algorithm, there is also a major conceptual difference in handling
missing symptoms across the algorithms. Missing symptoms could arise
from different stages of the data collection process. For example, the
respondent may not know whether certain symptoms existed or may refuse
to answer a question. From a statistical point of view, knowing that a
symptom does not exist provides some information to the possible cause
assignment, while a missing symptom does not. Although in theory, most
of the VA algorithms could benefit from distinguishing `missing' from
`absence', InSilicoVA is the only algorithm that has been implemented to
acknowledge missing data. Missing indicators are assumed to be
equivalent to `absence' in InterVA, NBC, and Tariff.

\hypertarget{data-preparation}{%
\section{Data preparation}\label{data-preparation}}

In the \CRANpkg{openVA} package, we consider two main types of
standardized questionnaire: the WHO instrument and the IHME
questionnaire. In this section, we focus on describing these two data
formats and tools to clean and convert data. Pre-processing the raw data
collected from the survey instrument (usually with Open Data Toolkit) is
usually performed with additional packages and software outside of the
analysis pipeline in R. We briefly mention software for data
pre-processing towards the end of this paper.

\hypertarget{the-who-standard-format}{%
\subsection{The WHO standard format}\label{the-who-standard-format}}

For users familiar with InterVA software and the associated data
processing steps, the standard input format from the WHO 2012 and 2016
instruments is usually well understood. For the 2012 instrument, the
data expected by the InterVA-4 software are organized into a data frame
where each row represents one death and the corresponding VA information
is contained in \(246\) fields, starting from the first item being the
ID of the death. The \(245\) items following the ID each represent one
binary variable of symptom/indicator, where `presence' is coded by `Y',
and `absence' is coded by an empty cell.

To accommodate updates for the WHO 2016 instrument
\citep{dambruoso2017}, the InterVA-5 software accepts a data frame with
\(354\) columns that include \(353\) columns of symptom/indicators
followed by an additional column for the record ID. It should be noted
that the R package \CRANpkg{InterVA5} retains the format with the record
ID residing in the first column. Another important update with InterVA-5
is that it acknowledges the difference between ``Yes'' and ``No'' (or
``Y/y'' and ``N/n'', which is different from the coding scheme in
InterVA-4), both of which are processed as relevant responses, while all
other responses are treated as missing values and ignored. With respect
to the list of causes of death, InterVA-5 utilizes the WHO 2016 COD
categories, which is nearly identical to the WHO 2012 COD categories
(used by InterVA-4) except that hemorrhagic fever and dengue fever are
two separate categories in the 2016 COD categories.

The same input format is inherited by the \CRANpkg{openVA} package,
except for one modification. We further distinguish `missing' and
`absence' in the input data frame explicitly. We highly recommend that
users pre-process all the raw data so that a `missing' value in the data
spreadsheet is coded as a `.' (following the Stata practice familiar to
many VA practitioners), and an `absence' value is indicated by an empty
cell, as in the standard InterVA-4 software. For WHO 2016 data, both `.'
and `-' (following the default coding scheme of InterVA-5 software) are
interpreted as missing values. For methods other than InSilicoVA,
`missing' and `absence' will be considered the same internally and thus
will not introduce a compatibility problem.

\hypertarget{the-phmrc-format}{%
\subsection{The PHMRC format}\label{the-phmrc-format}}

The Population Health Metrics Research Consortium (PHMRC) gold standard
VA data \citep{murray2011population} consist of three datasets
corresponding to adult, child, and neonatal deaths, respectively. All
deaths occurred in health facilities and gold-standard causes are
determined based on laboratory, pathology and medical imaging findings.
These datasets can be downloaded directly using the link returned by the
function \texttt{getPHMRC\_url()}. For example, we can read the adult VA
dataset using the following command.

\begin{Schunk}
\begin{Sinput}
PHMRC_adult <- read.csv(getPHMRC_url("adult"))
\end{Sinput}
\end{Schunk}

Although the data are publicly accessible, a major practical challenge
for studies involving the PHMRC gold standard dataset is that the
pre-processing steps described from the relevant publications are not
clear enough nor easy to implement. The \CRANpkg{openVA} package
internally cleans up the PHMRC gold standard data when calling the
\texttt{codeVA()} function on the PHMRC data. The procedure follows the
steps described in the supplement material of \citet{insilico}. Notice
that the original PHMRC data are useful for comparing and validating new
methods, as well as for using as training data, but the cleaning
functions only require that the columns are exactly the same as the
PHMRC gold standard datasets, so they could also be used for new data
that are pre-processed into the same format.

\hypertarget{customized-format}{%
\subsection{Customized format}\label{customized-format}}

In addition to the two standard questionnaires discussed previously,
researchers might also be interested in including customized dichotomous
symptoms in their analysis. The \CRANpkg{openVA} package also supports
customized inputs as long as they are dichotomous. In such case, neither
the built-in conditional probability matrix of InterVA nor the PHMRC
gold standard dataset could be used to learn the relationship between
training and testing data, thus different training data with known
causes of death are necessary for all three algorithms. The
\texttt{ConvertData()} function can be used to convert data with
customized coding schemes into the format recognized by the
\CRANpkg{openVA} package.

Finally, we note that the \CRANpkg{openVA} package currently does not
reformat data from one standardized questionnaire to another. This is
because mapping the symptoms collected from one questionnaire to those
collected by another questionnaire inevitably creates loss of
information. Such mapping tasks can be useful for some applications. For
example, a full mapping of a PHMRC dataset into the WHO format enables
the use of physician provided conditional probabilities included in the
InterVA software on data collected by PHMRC questionnaires. This remains
as an important feature to be added to the package in the future.

\hypertarget{fitting-va-cause-of-death-assignment-models}{%
\section{Fitting VA cause-of-death assignment
models}\label{fitting-va-cause-of-death-assignment-models}}

In this section, we demonstrate the model fitting process in the
\CRANpkg{openVA} package using two datasets: (1) a random sample of
\(1,000\) deaths from the ALPHA network without cause-of-death labels
collected with the WHO 2012 instrument, and (2) the adult VA records in
the PHMRC gold standard data, with the \(1,554\) records from Andhra
Pradesh, India used as a testing set and the rest used as a training
dataset. In the first case without gold standard training data, only
InterVA and InSilicoVA can be fitted. All four methods can be fitted in
the second case.

\hypertarget{modeling-data-collected-with-who-2012-questionnaire}{%
\subsection{Modeling data collected with WHO 2012
questionnaire}\label{modeling-data-collected-with-who-2012-questionnaire}}

The randomly sampled VA records from the ALPHA network sites are already
included in the openVA package as a dataset \texttt{RandomVA1} and can
be loaded directly.

\begin{Schunk}
\begin{Sinput}
data(RandomVA1)
dim(RandomVA1)
\end{Sinput}
\begin{Soutput}
#> [1] 1000  246
\end{Soutput}
\begin{Sinput}
head(RandomVA1[, 1:10])
\end{Sinput}
\begin{Soutput}
#>   ID elder midage adult child under5 infant neonate male female
#> 1 d1     Y                                             Y       
#> 2 d2     Y                                                    Y
#> 3 d3            Y                                      Y       
#> 4 d4                  Y                                       Y
#> 5 d5                  Y                                Y       
#> 6 d6                  Y                                       Y
\end{Soutput}
\end{Schunk}

The \texttt{codeVA()} function provides a standardized syntax to fit
different VA models. Internally, the \texttt{codeVA()} function
organizes the input data according to the specified data type, checks
for incompatibility of the data and specified model, and calls the
corresponding model fitting functions. It returns a classed object of
the specified model class. In this example, we use version 4.03 of the
InterVA software, which is the latest release of the original software
compatible with the WHO 2012 instrument. Any additional model-specific
parameters can be passed through the arguments of \texttt{codeVA()}.
Here we specify the HIV and malaria prevalence levels required by the
InterVA model to be `high'. Guidelines on how to set these parameters
can be found in \citet{byass2012strengthening}.

\begin{Schunk}
\begin{Sinput}
fit_inter_who <- codeVA(data = RandomVA1, data.type = "WHO2012", 
                        model = "InterVA", version = "4.03", 
                        HIV = "h", Malaria = "h")
\end{Sinput}
\end{Schunk}

\begin{Schunk}
\begin{Sinput}
summary(fit_inter_who) 
\end{Sinput}
\begin{Soutput}
#> InterVA-4 fitted on 1000 deaths
#> CSMF calculated using reported causes by InterVA-4 only
#> The remaining probabilities are assigned to 'Undetermined'
#> 
#> Top 5 CSMFs:
#>  cause                     likelihood
#>  Undetermined              0.154     
#>  HIV/AIDS related death    0.122     
#>  Stroke                    0.072     
#>  Reproductive neoplasms MF 0.058     
#>  Pulmonary tuberculosis    0.055
\end{Soutput}
\end{Schunk}

We can implement InSilicoVA method with similar syntax. We use the
default parameters and run the MCMC for \(10,000\) iterations. Setting
the \texttt{auto.length} argument to FALSE specifies that the algorithm
does not automatically increase the length of the chain when convergence
failed. The InSilicoVA algorithm is implemented using a
Metropolis-Hastings within Gibbs sampler. The acceptance rate is printed
as part of the message as the model samples from the posterior
distribution.

\begin{Schunk}
\begin{Sinput}
fit_ins_who <- codeVA(RandomVA1, data.type = "WHO2012", model = "InSilicoVA",
                    Nsim = 10000, auto.length = FALSE)
\end{Sinput}
\end{Schunk}

\begin{Schunk}
\begin{Sinput}
summary(fit_ins_who) 
\end{Sinput}
\begin{Soutput}
#> InSilicoVA Call: 
#> 1000 death processed
#> 10000 iterations performed, with first 5000 iterations discarded
#>  250 iterations saved after thinning
#> Fitted with re-estimated conditional probability level table
#> Data consistency check performed as in InterVA4
#> 
#> Top 10 CSMFs:
#>                                    Mean Std.Error Lower Median Upper
#> Other and unspecified infect dis  0.266    0.0168 0.235  0.265 0.301
#> HIV/AIDS related death            0.102    0.0091 0.085  0.102 0.119
#> Renal failure                     0.101    0.0108 0.084  0.101 0.123
#> Other and unspecified neoplasms   0.062    0.0089 0.046  0.061 0.080
#> Other and unspecified cardiac dis 0.058    0.0076 0.044  0.058 0.075
#> Digestive neoplasms               0.050    0.0077 0.033  0.050 0.065
#> Acute resp infect incl pneumonia  0.048    0.0073 0.034  0.049 0.063
#> Pulmonary tuberculosis            0.039    0.0068 0.025  0.039 0.054
#> Stroke                            0.038    0.0061 0.027  0.038 0.052
#> Other and unspecified NCD         0.034    0.0089 0.018  0.034 0.052
\end{Soutput}
\end{Schunk}

\hypertarget{modeling-the-phmrc-data}{%
\subsection{Modeling the PHMRC data}\label{modeling-the-phmrc-data}}

In the second example, we consider a prediction task using the PHMRC
adult dataset. We first load the complete PHMRC adult dataset from its
on-line repository, and organize it into training and test datasets. We
treat all deaths from Andhra Pradesh, India as the test dataset.

\begin{Schunk}
\begin{Sinput}
PHMRC_adult <- read.csv(getPHMRC_url("adult"))
is.test <- which(PHMRC_adult$site == "AP")
test <- PHMRC_adult[is.test, ]
train <- PHMRC_adult[-is.test, ]
dim(test)
\end{Sinput}
\begin{Soutput}
#> [1] 1554  946
\end{Soutput}
\begin{Sinput}
dim(train)
\end{Sinput}
\begin{Soutput}
#> [1] 6287  946
\end{Soutput}
\end{Schunk}

In order to fit the models on the PHMRC data, we specify
\texttt{data.type\ =\ "PHMRC"} and \texttt{phmrc.type\ =\ "adult"} to
indicate the data input is collected using the PHMRC adult
questionnaire. We also specify the column of the causes-of-death label
in the training data. The rest of the syntax is similar to the previous
example.

When the input consists of both training and testing data, the InterVA
and InSilicoVA algorithms estimate the conditional probabilities of
symptoms using the training data, instead of using the built-in values.
In such case, the \texttt{version} argument for the InterVA algorithm is
suppressed. There are several ways to map the conditional probabilities
of symptoms given causes in the training dataset to a letter grade
system, specified by the \texttt{convert.type} argument. The
\texttt{convert.type\ =\ "quantile"} performs the mapping so that the
percentile of each rank stays the same as the original \(P_{s|c}\)
matrix in InterVA software. Alternatively we can also use the original
fixed values of translation, and assign letter grades closest to each
entry in \(\hat{P}_{s|c}\). This conversion is specified by
\texttt{convert.type\ =\ "fixed"}, and is more closely aligned to the
original InterVA and InSilicoVA setting. Finally, we can also directly
use the values in the \(\hat{P}_{s|c}\) without converting them to ranks
and re-estimating the values associated with each rank. This can be
specified by \texttt{convert.type\ =\ "empirical"}. In this
demonstration, we assume the fixed value conversion.

\begin{Schunk}
\begin{Sinput}
fit_inter <- codeVA(data = test, data.type = "PHMRC", model = "InterVA", 
                     data.train = train, causes.train = "gs_text34", 
                     phmrc.type = "adult", convert.type = "fixed")
\end{Sinput}
\end{Schunk}

\begin{Schunk}
\begin{Sinput}
fit_ins <- codeVA(data = test, data.type = "PHMRC", model = "InSilicoVA",
                    data.train = train, causes.train = "gs_text34", 
                    phmrc.type = "adult", convert.type = "fixed", 
                    Nsim=10000, auto.length = FALSE)
\end{Sinput}
\end{Schunk}

The NBC and Tariff method can be fit using similar syntax.

\begin{Schunk}
\begin{Sinput}
fit_nbc <- codeVA(data = test, data.type = "PHMRC", model = "NBC", 
                   data.train = train, causes.train = "gs_text34", 
                   phmrc.type = "adult")
\end{Sinput}
\end{Schunk}

\begin{Schunk}
\begin{Sinput}
fit_tariff <- codeVA(data = test, data.type = "PHMRC", model = "Tariff",
                     data.train = train, causes.train = "gs_text34", 
                     phmrc.type = "adult")
\end{Sinput}
\end{Schunk}

Notice that we do not need to transform the PHMRC data manually. Data
transformations are performed automatically within the \texttt{codeVA()}
function.

\hypertarget{summarizing-results}{%
\section{Summarizing results}\label{summarizing-results}}

In this section we demonstrate how to summarize results, extract output,
and visualize and compare fitted results. All the fitted object returned
by \texttt{codeVA()} are S3 objects, for which a readable summary of
model results can be obtained with the \texttt{summary()} function as
shown in the previous section. In addition, several other metrics are
commonly used to evaluate and compare VA algorithms at either the
population or individual levels. In the rest of this section, we show
how to easily calculate and visualize some of these metrics with the
\CRANpkg{openVA} package.

\hypertarget{csmf-accuracy}{%
\subsection{CSMF Accuracy}\label{csmf-accuracy}}

We can extract the CSMFs directly using the \texttt{getCSMF()} function.
The function returns a vector of the point estimates of the CSMFs, or a
matrix of posterior summaries of the CSMF for the InSilicoVA algorithm.

\begin{Schunk}
\begin{Sinput}
csmf_inter <- getCSMF(fit_inter)
csmf_ins <- getCSMF(fit_ins)
csmf_nbc <- getCSMF(fit_nbc)
csmf_tariff <- getCSMF(fit_tariff)
\end{Sinput}
\end{Schunk}

One commonly used metric to evaluate the CSMF estimates is the so-called
CSMF accuracy, defined as \[
CSMF_{acc} = 1 - \frac{\sum_j^C CSMF_i - CSMF_j^{(true)}}{2(1 - \min CSMF^{(true)})}
\] The CSMF accuracy can be readily computed using functions in
\CRANpkg{openVA} as the codes below shows.

\begin{Schunk}
\begin{Sinput}
csmf_true <- table(c(test$gs_text34, unique(PHMRC_adult$gs_text34))) - 1
csmf_true <- csmf_true / sum(csmf_true)
c(getCSMF_accuracy(csmf_inter, csmf_true, undet = "Undetermined"), 
  getCSMF_accuracy(csmf_ins[, "Mean"], csmf_true), 
  getCSMF_accuracy(csmf_nbc, csmf_true), 
  getCSMF_accuracy(csmf_tariff, csmf_true))
\end{Sinput}
\begin{Soutput}
#> [1] 0.53 0.74 0.77 0.68
\end{Soutput}
\end{Schunk}

We use the empirical distribution in the test data to calculate the true
CSMF distribution, i.e.,
\(CSMF_j^{(true)} = \frac{1}{n}\sum_{i=1}^n \bm{1}_{y_i = j}\). Then we
evaluate the CSMF accuracy using the \texttt{getCSMF\_accuracy()}
function. As discussed previously, the default CSMF calculation is
slightly different for diCfferent methods. For example, the InterVA
algorithm creates the additional category of \texttt{Undetermined} by
default, which is not in the true CSMF categories and needs to be
specified. The creation of the undetermined category can also be
suppressed by \texttt{interVA.rule\ =\ FALSE} in the \texttt{getCSMF()}
function call. For the InSilicoVA algorithm, we use the posterior mean
to calculate the point estimates of the CSMF accuracy.

\hypertarget{individual-cod-summary}{%
\subsection{Individual COD summary}\label{individual-cod-summary}}

At the individual level, we can extract the most likely cause-of-death
assignment from the fitted object using the \texttt{getTopCOD()}
function.

\begin{Schunk}
\begin{Sinput}
cod_inter <- getTopCOD(fit_inter)
cod_ins <- getTopCOD(fit_ins)
cod_nbc <- getTopCOD(fit_nbc)
cod_tariff <- getTopCOD(fit_tariff)
\end{Sinput}
\end{Schunk}

With the most likely COD assignment, other types of metrics based on
individual COD assignment accuracy can be similarly constructed by
users. The summary methods can also be called for each death ID. For
example, using the Tariff method, we can extract the fitted rankings of
causes for the death with ID 6288 by

\begin{Schunk}
\begin{Sinput}
summary(fit_inter, id = "6288")
\end{Sinput}
\begin{Soutput}
#> InterVA-4 fitted top 5 causes for death ID: 6288
#> 
#>  Cause                     Likelihood
#>  Stroke                    0.509     
#>  Pneumonia                 0.318     
#>  COPD                      0.081     
#>  Other Infectious Diseases 0.064     
#>  Renal Failure             0.013
\end{Soutput}
\end{Schunk}

The \texttt{summary()} function for InSilcoVA does not provide
uncertainty estimates for individual COD assignments by default. This is
because, in practice, the calculation of individual posterior
probabilities of COD distribution can be memory-intensive when the
dataset is large. To obtain individual-level uncertainty measurements,
we can either run the MCMC chain with the additional argument
\texttt{indiv.CI\ =\ 0.95} when calling \texttt{codeVA()}, or update the
fitted object directly with the saved posterior draws.

\begin{Schunk}
\begin{Sinput}
fit_ins <- updateIndiv(fit_ins, CI = 0.95)
summary(fit_ins, id = "6288")
\end{Sinput}
\begin{Soutput}
#> InSilicoVA fitted top  causes for death ID: 6288
#> Credible intervals shown: 95%
#>                               Mean  Lower Median  Upper
#> Stroke                      0.5043 0.3485 0.5083 0.6361
#> Pneumonia                   0.4116 0.2615 0.4083 0.5834
#> Other Infectious Diseases   0.0660 0.0411 0.0642 0.0966
#> Epilepsy                    0.0099 0.0064 0.0097 0.0142
#> COPD                        0.0053 0.0031 0.0052 0.0079
#> Malaria                     0.0007 0.0005 0.0007 0.0011
#> Diabetes                    0.0005 0.0003 0.0005 0.0009
#> Acute Myocardial Infarction 0.0004 0.0003 0.0004 0.0006
#> Falls                       0.0004 0.0001 0.0004 0.0013
#> Renal Failure               0.0004 0.0002 0.0003 0.0005
\end{Soutput}
\end{Schunk}

For \(N\) deaths, \(C\) causes, the posterior mean of individual COD
distributions returned by the InSilicoVA model, along with median and
with credible intervals can be represented by a
\((N \times C \times 4)\)-dimensional array. The function
\texttt{getIndivProb()} extracts this summary in the form of a list of
\(4\) matrices of dimension \(N\) by \(C\), which can then be saved to
other formats to facilitate further analysis. For other methods, the
point estimates of individual COD distribution are returned as the \(N\)
by \(C\) matrix.

\begin{Schunk}
\begin{Sinput}
fit_prob <- getIndivProb(fit_inter)
dim(fit_prob)
\end{Sinput}
\begin{Soutput}
#> [1] 1554   34
\end{Soutput}
\end{Schunk}

\hypertarget{visualization}{%
\subsection{Visualization}\label{visualization}}

The previous sections discuss how results could be extracted and
examined in R. In this subsection, we show some visualization tools
provided in the \CRANpkg{openVA} package for presenting these results.
The fitted CSMFs for the top causes can be easily visualized by the
\texttt{plotVA()} function. The default graph components are specific to
each algorithm and individual package implementations, with options for
further customization. For example, Figure\ref{fig:vis-1} shows the
estimated CSMF from the InterVA algorithm in the PHMRC data example.

\begin{Schunk}
\begin{Sinput}
plotVA(fit_inter, title = "InterVA")
\end{Sinput}
\begin{figure}[!h]

{\centering \includegraphics[width=0.8\linewidth,]{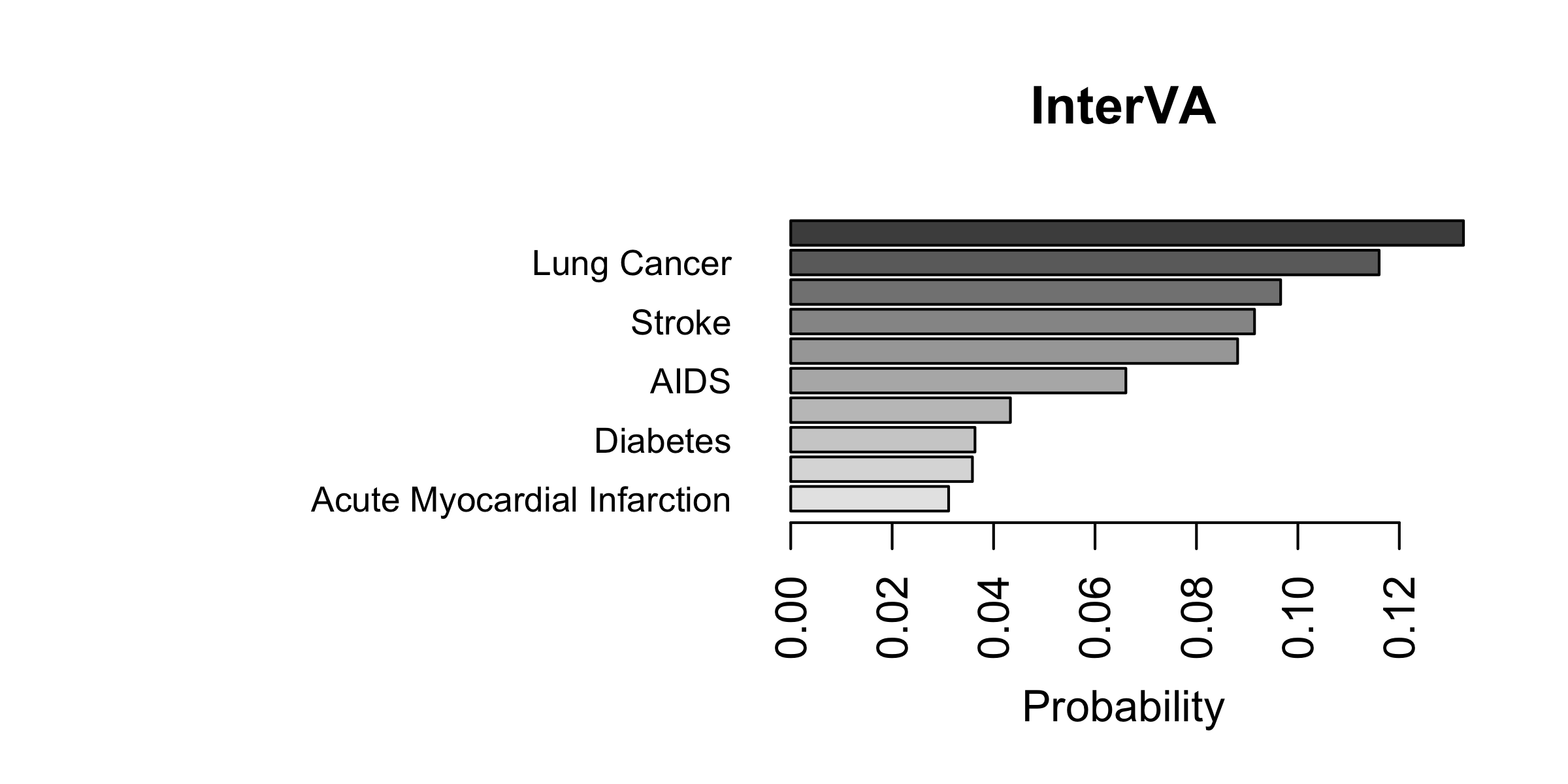} 

}

\caption[Top 10 CSMFs estimated by InterVA]{Top 10 CSMFs estimated by InterVA.}\label{fig:vis-1}
\end{figure}
\end{Schunk}

The CSMFs can also be aggregated for easier visualization of groups of
causes. For the InterVA-4 cause list, we included an example grouping
built into the package, so the aggregated CSMFs can be compared
directly. In practice, the grouping of causes of deaths often needs to
be determined according to context and the research question of
interest. Changing the grouping can be easily achieved by modifying the
\texttt{grouping} argument in the \texttt{stackplotVA()} function. For
example, to facilitate the new category of \texttt{Undetermined}
returned by InterVA, we first modify the grouping matrix to include it
as a new cause and visualize the aggregated CSMF estimates in Figure
\ref{fig:vis-2}.

\begin{Schunk}
\begin{Sinput}
data(SampleCategory)
grouping <- SampleCategory
grouping[,1] <- as.character(grouping[,1])
grouping <- rbind(grouping, c("Undetermined", "Undetermined"))
compare <- list(InterVA4 = fit_inter_who,
                InSilicoVA = fit_ins_who)
stackplotVA(compare, xlab = "", angle = 0,  grouping = grouping)
\end{Sinput}
\begin{figure}[!h]

{\centering \includegraphics[width=.7\linewidth,]{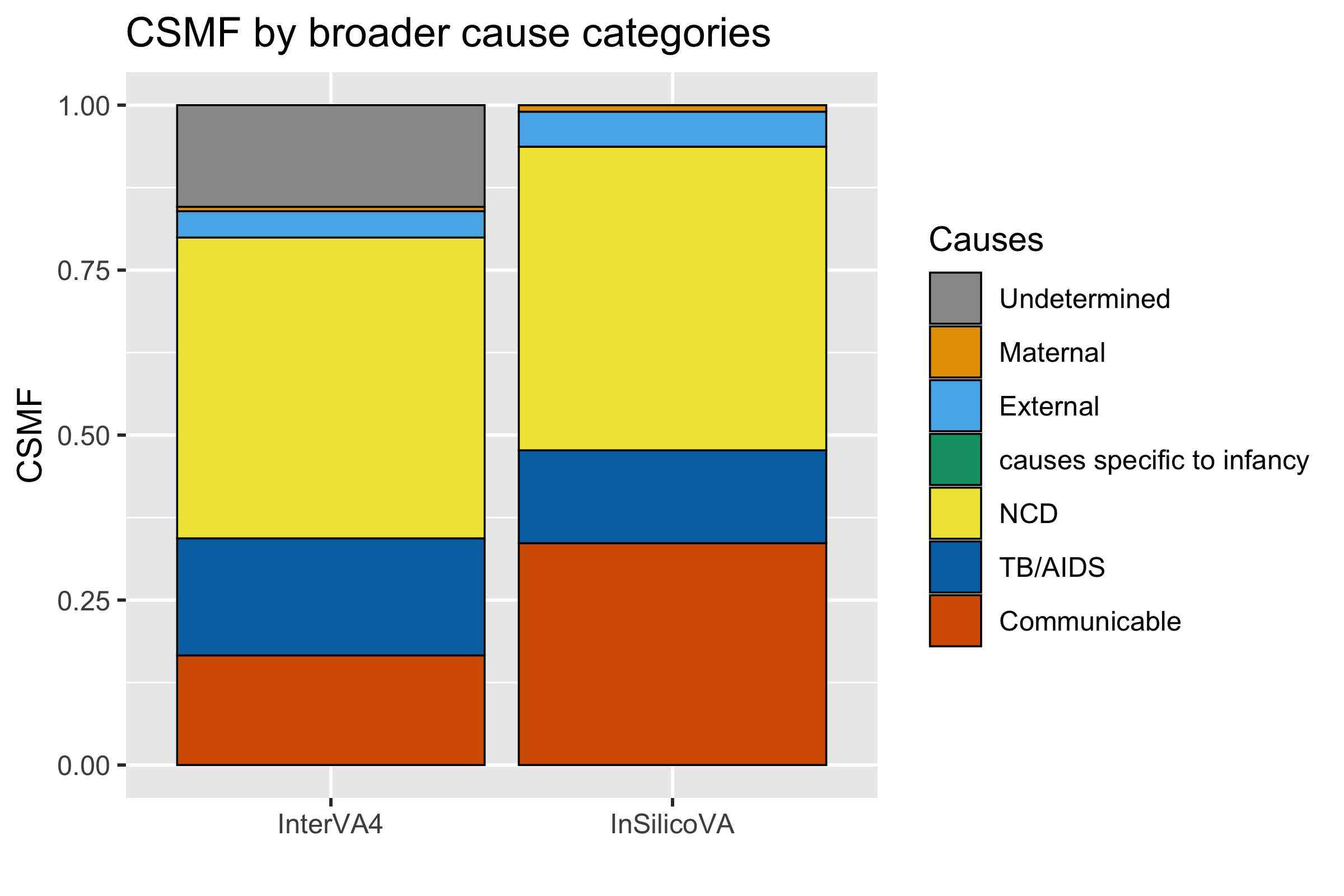} 

}

\caption[Estimated aggregated CSMFs for InterVA-4 and InSilicoVA, adding undetermined category]{Estimated aggregated CSMFs for InterVA-4 and InSilicoVA, adding undetermined category.}\label{fig:vis-2}
\end{figure}
\end{Schunk}

The ordering of the stacked bars can also be changed to reflect the
structures within the aggregated causes, as demonstrated in Figure
\ref{fig:vis-3}.

\begin{Schunk}
\begin{Sinput}
group_order <- c("TB/AIDS",  "Communicable", "NCD", "External", "Maternal",
            "causes specific to infancy", "Undetermined") 
stackplotVA(compare, xlab = "", angle = 0, grouping = grouping, 
            group_order = group_order)
\end{Sinput}
\begin{figure}[!h]

{\centering \includegraphics[width=.7\linewidth,]{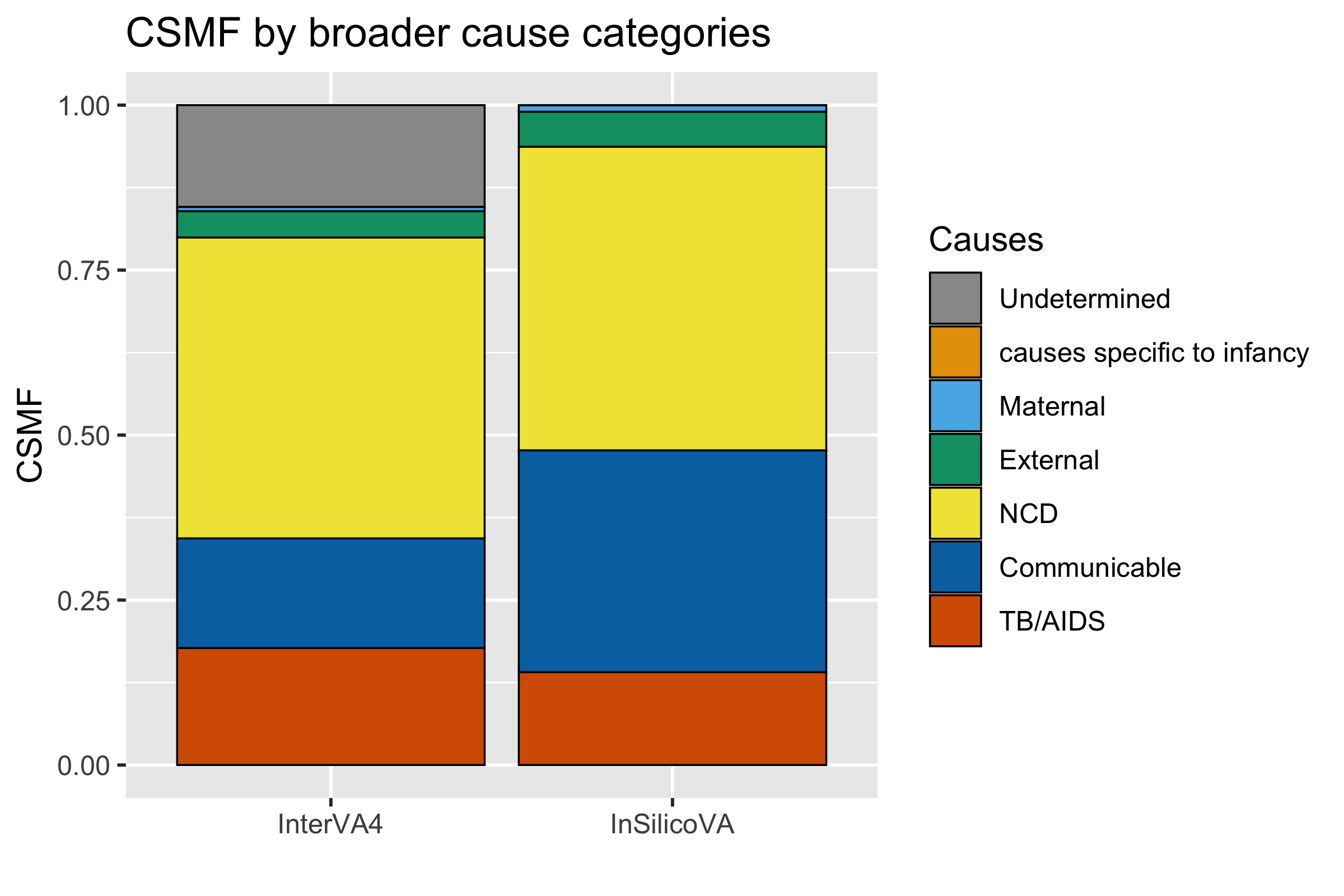} 

}

\caption[Estimated aggregated CSMFs for InterVA-4 and InSilicoVA, with the causes reordered]{Estimated aggregated CSMFs for InterVA-4 and InSilicoVA, with the causes reordered.}\label{fig:vis-3}
\end{figure}
\end{Schunk}

\hypertarget{incorporating-additional-information}{%
\section{Incorporating additional
information}\label{incorporating-additional-information}}

Among the VA methods discussed in this paper, the InSilicoVA algorithm
\citep{insilico} allows for more flexible modifications to the Bayesian
hierarchical model structure when additional information is available.
In this section, we illustrate two features unique to the InSilicoVA
method: jointly estimating CSMFs from multiple populations, and
incorporating partial and potentially noisy physician codings into the
algorithm.

\hypertarget{sub-population-specific-csmfs}{%
\subsection{Sub-population specific
CSMFs}\label{sub-population-specific-csmfs}}

In practice researchers may want to estimate and compare CSMFs for
different regions, time periods, or demographic groups in the
population. Running separate models on subsets of data can be
inefficient and does not allow parameter estimation to borrow
information across different groups. The generative framework adopted by
InSilicoVA allows the specification of sub-populations in analyzing VA
data. Consider an input dataset with \(G\) different sub-populations. We
can estimate different CSMFs \(\pi^{(g)}\) for \(g = 1, ..., G\) for
each sub-population, while assuming the same conditional probability
matrix, \(P_{s|c}\) and other hyperpriors. As an example, we show how to
estimate different CSMFs for sub-populations specified by sex and age
groups, using a randomly sampled ALPHA dataset with additional columns
specifying the sub-population each death belongs to.

\begin{Schunk}
\begin{Sinput}
data(RandomVA2)
head(RandomVA2[, 244:248])
\end{Sinput}
\begin{Soutput}
#>   stradm smobph scosts   sex age
#> 1      .      .      .   Men 60+
#> 2      .      .      . Women 60-
#> 3      .      .      . Women 60-
#> 4      .      .      . Women 60+
#> 5      .      .      . Women 60-
#> 6      .      .      . Women 60-
\end{Soutput}
\end{Schunk}

Then we can fit the model with one or multiple additional columns
specifying sub-population membership for each observation.

\begin{Schunk}
\begin{Sinput}
fit_sub <- codeVA(RandomVA2, model = "InSilicoVA", 
              subpop = list("sex", "age"),  indiv.CI = 0.95,
              Nsim = 10000, auto.length = FALSE)
\end{Sinput}
\end{Schunk}

Functions discussed in the previous sections work in the same way for
the fitted object with multiple sub-populations. Additional
visualization tools are also available. Figure \ref{fig:ins3} plots the
CSMFs for two sub-populations on the same plot by specify
\texttt{type\ =\ "compare"}.

\begin{Schunk}
\begin{Sinput}
plotVA(fit_sub, type = "compare", title = "Comparing CSMFs", top = 3)
\end{Sinput}
\begin{figure}[!h]

{\centering \includegraphics[width=0.9\linewidth,]{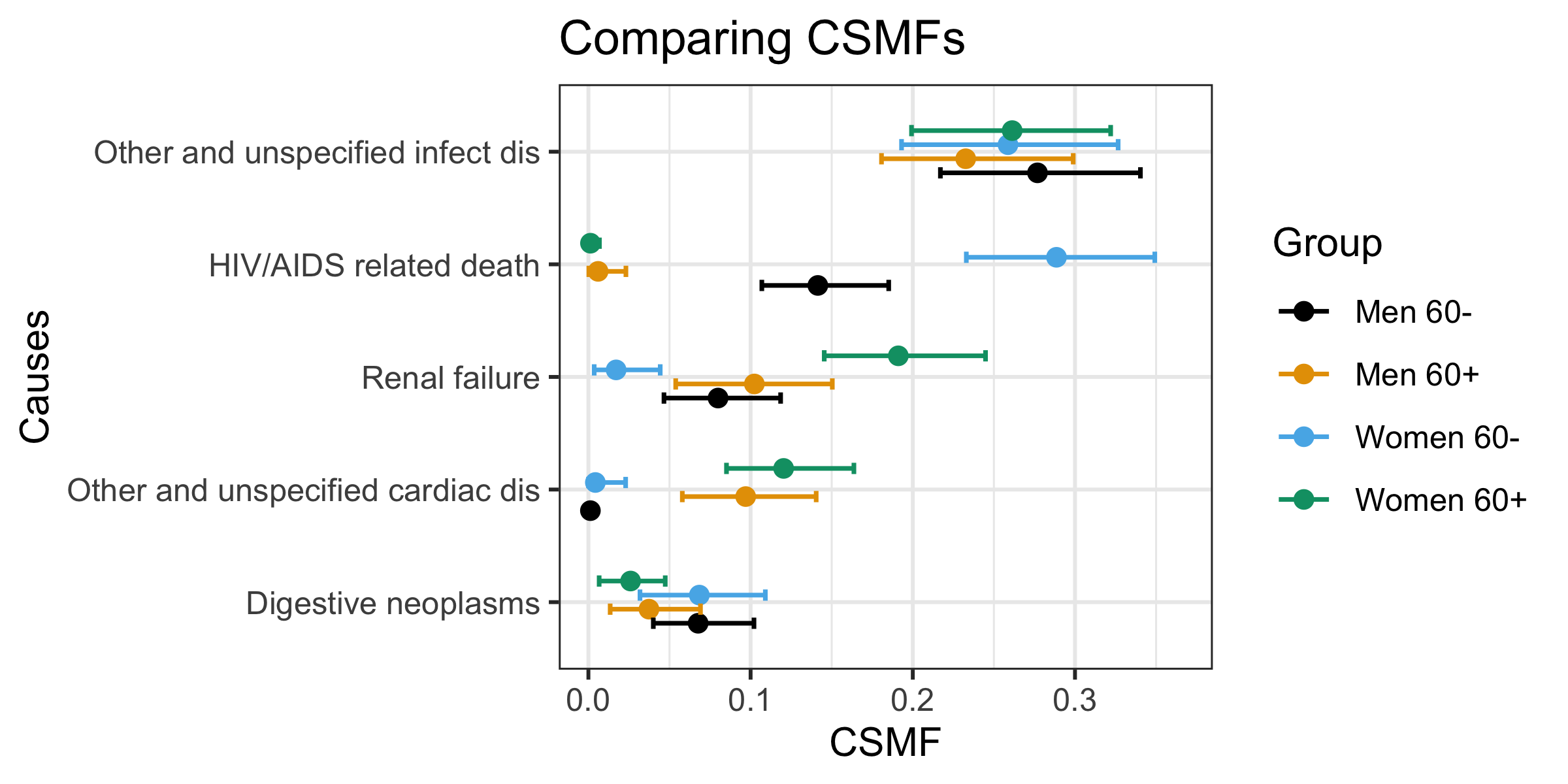} 

}

\caption[Estimated CSMFs for different sub-populations]{Estimated CSMFs for different sub-populations. The points indicate posterior means of the CSMF and the error bars indicate 95\% credible intervals of the CSMF.}\label{fig:ins3}
\end{figure}
\end{Schunk}

By default, the comparison plots will select all the CODs that are
included in the top \(K\) for each of the sub-populations. We can also
plot only subsets of them by specifying the causes of interest, as shown
in Figure \ref{fig:ins4}.

\begin{Schunk}
\begin{Sinput}
plotVA(fit_sub, type = "compare", title = "Comparing CSMFs",
                causelist = c("HIV/AIDS related death", 
                              "Pulmonary tuberculosis", 
                              "Other and unspecified infect dis", 
                              "Other and unspecified NCD"))
\end{Sinput}
\begin{figure}[!h]

{\centering \includegraphics[width=0.8\linewidth,]{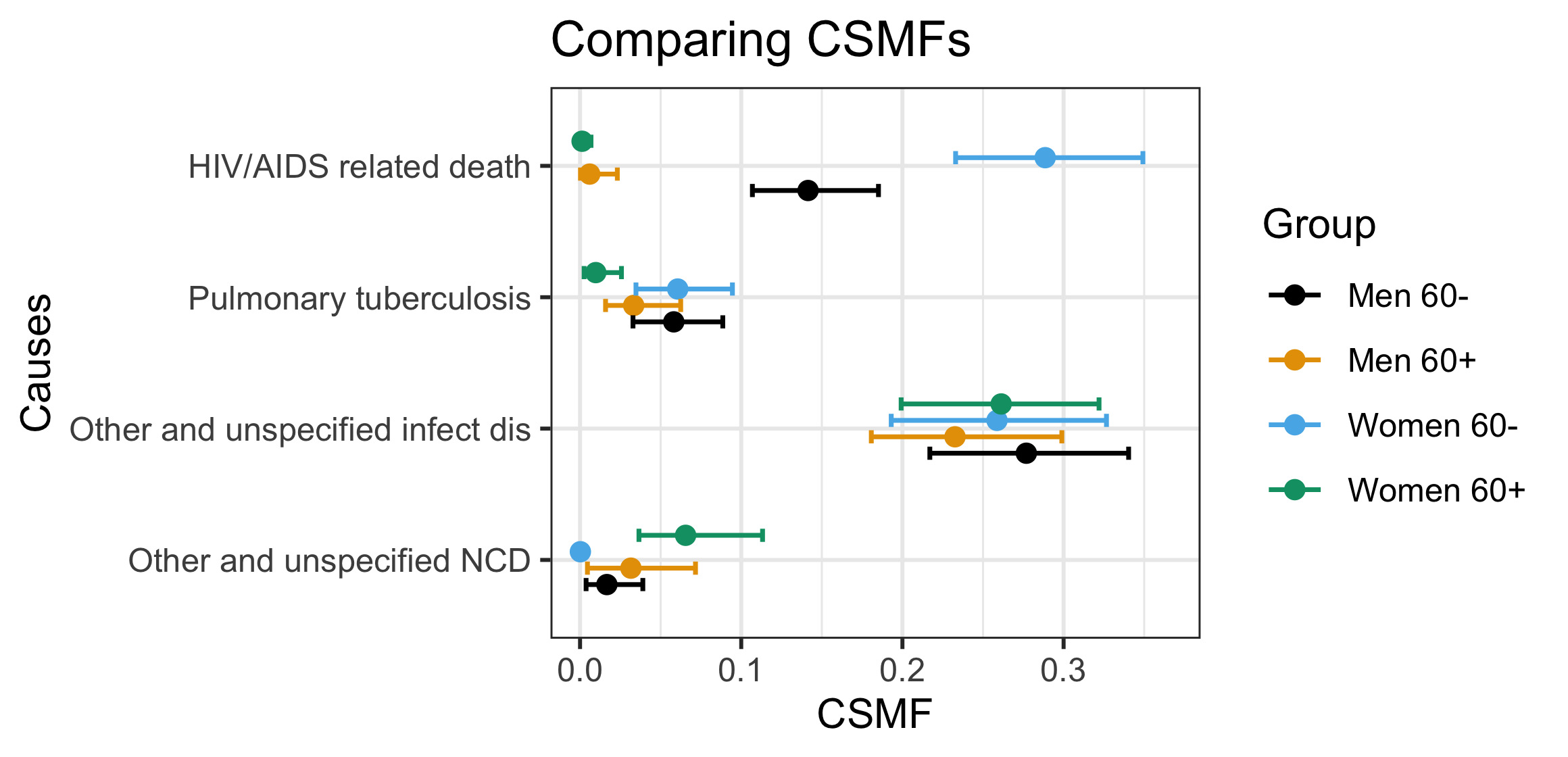} 

}

\caption[ Estimated fraction of deaths due to selected CODs for different sub-populations]{ Estimated fraction of deaths due to selected CODs for different sub-populations. The points indicate posterior means of the CSMF and the error bars indicate 95\% credible intervals of the CSMF.}\label{fig:ins4}
\end{figure}
\end{Schunk}

Figure \ref{fig:ins5} shows the visualization of the top CSMFs for a
chosen sub-population using the \texttt{which.sub} argument.

\begin{Schunk}
\begin{Sinput}
plotVA(fit_sub, which.sub = "Women 60-", title = "Women 60-")
\end{Sinput}
\begin{figure}[!h]

{\centering \includegraphics[width=0.8\linewidth,]{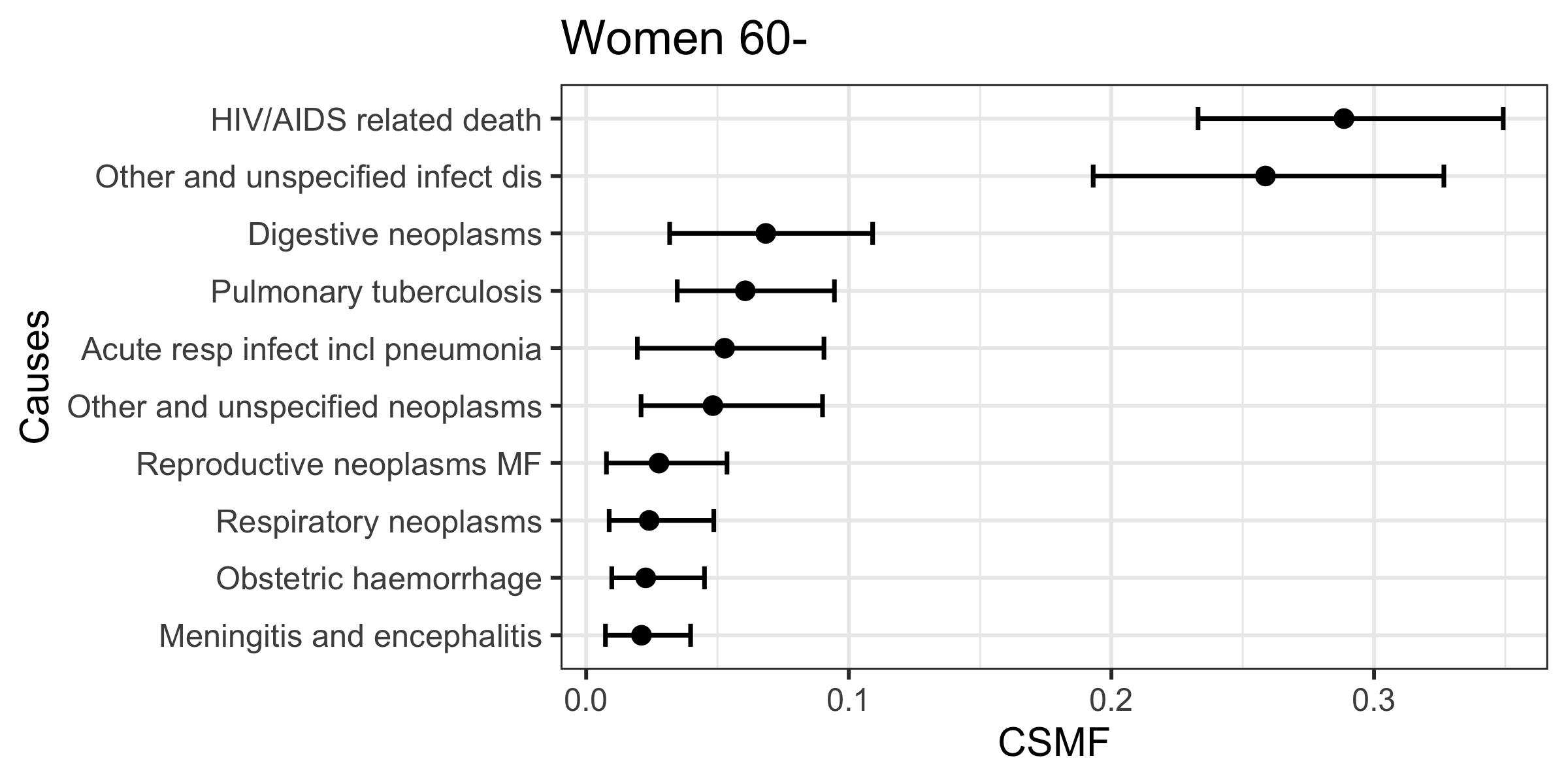} 

}

\caption[Top 10 CSMFs for a specified sub-population (women under $60$ years old)]{Top 10 CSMFs for a specified sub-population (women under $60$ years old).}\label{fig:ins5}
\end{figure}
\end{Schunk}

Similar to before, the \texttt{stackplot()} function can also be used to
compare different sub-populations in aggregated cause groups, as shown
in Figure \ref{fig:ins6}.

\begin{Schunk}
\begin{Sinput}
stackplotVA(fit_sub)
stackplotVA(fit_sub, type = "dodge")
\end{Sinput}
\begin{figure}[!h]

{\centering \includegraphics[width=0.49\linewidth,]{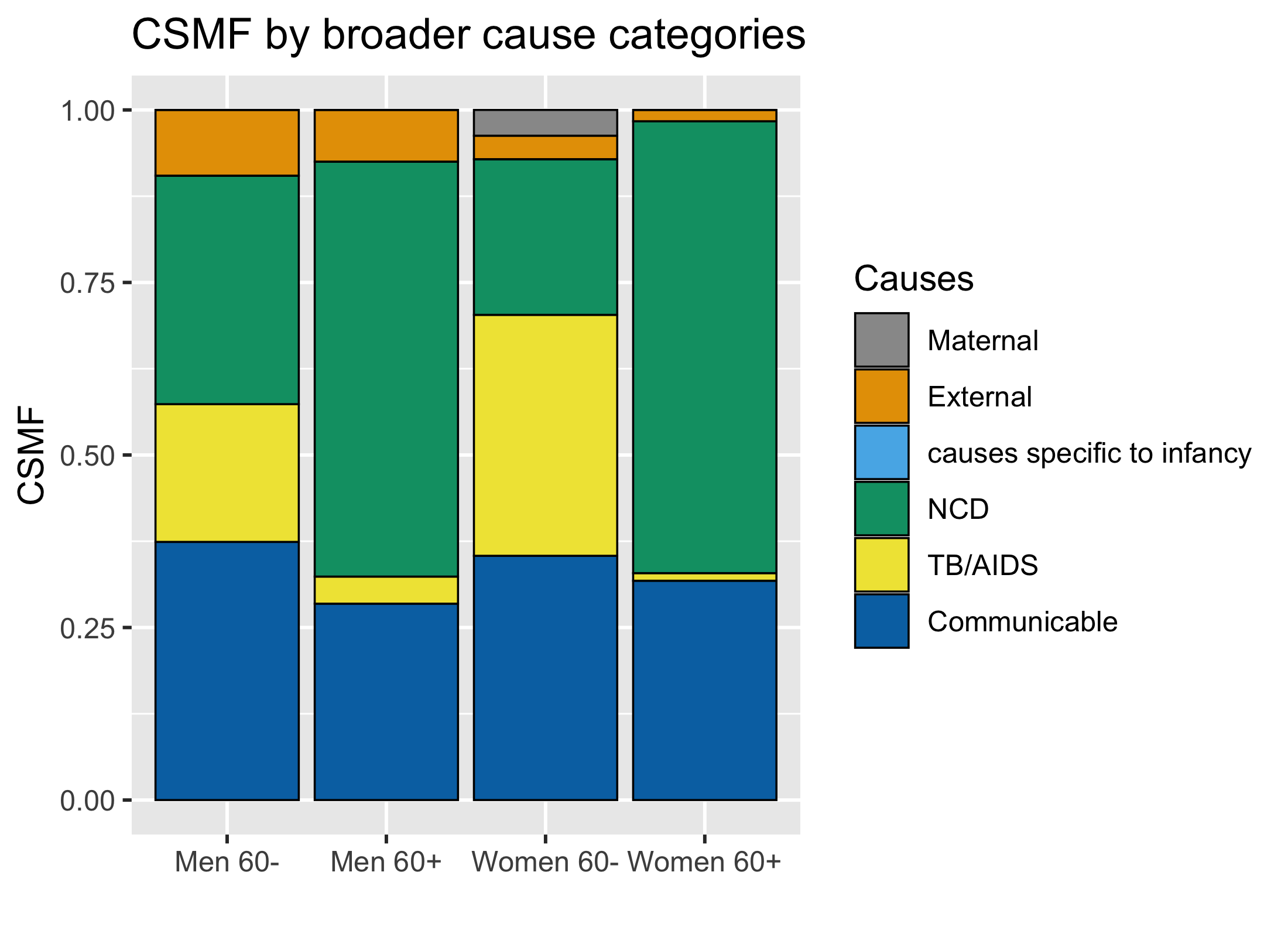} \includegraphics[width=0.49\linewidth,]{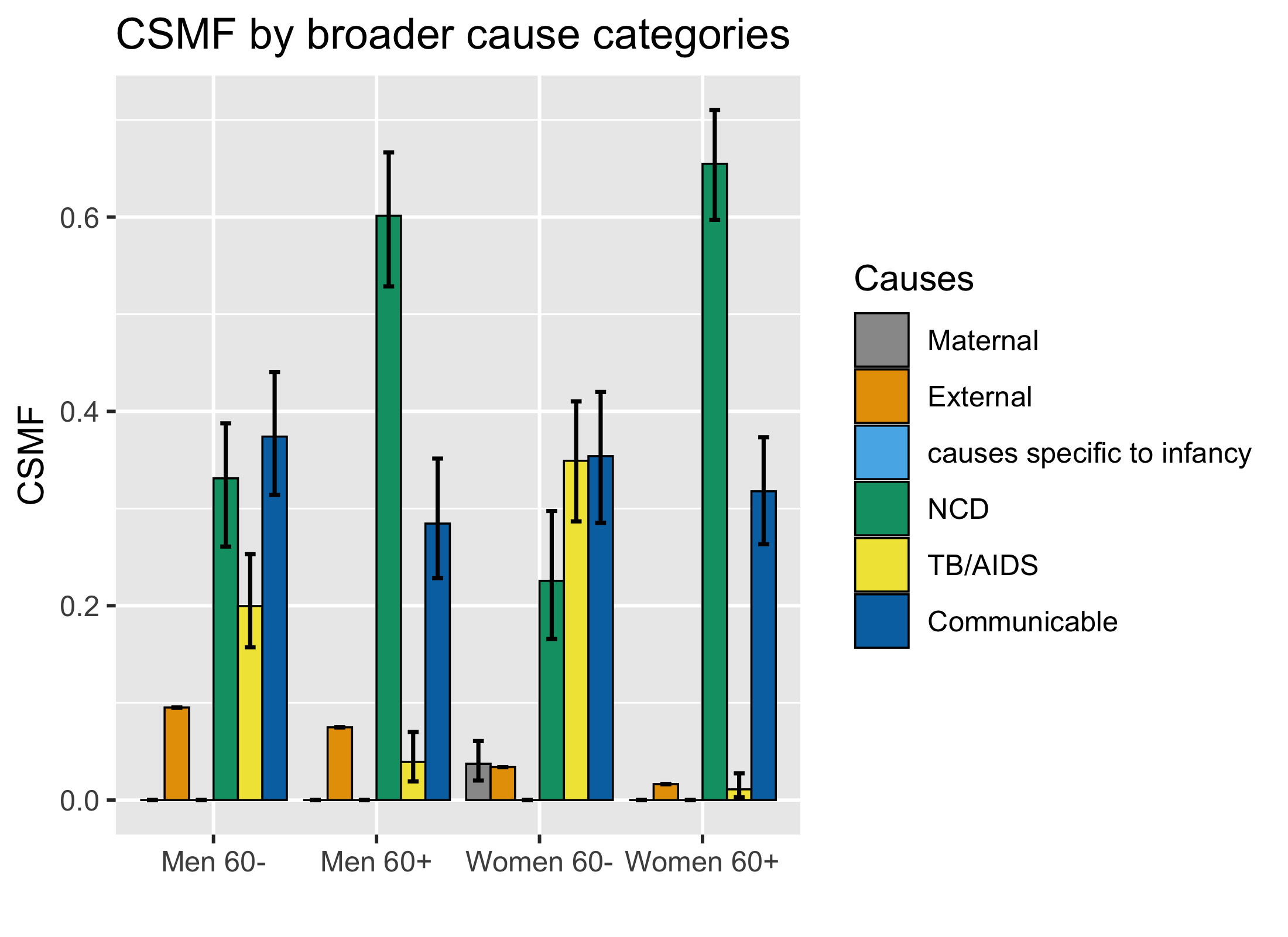} 

}

\caption[Aggregated CSMFs for four different sub-populations]{Aggregated CSMFs for four different sub-populations. Left: comparison using the stacked bar chart. Right: comparison using the bar chart arranged side to side. The height of the bars indicate posterior means of the CSMF and the error bars indicate 95\% credible intervals of the CSMF.}\label{fig:ins6}
\end{figure}
\end{Schunk}

\hypertarget{physician-coding}{%
\subsection{Physician coding}\label{physician-coding}}

When physician-coded causes of death are available for all or a subset
of the deaths, we can incorporate such information in the InSilicoVA
model. The physician-coded causes can be either the same as the CODs
used for the final algorithm, or consist of causes at a higher level of
aggregation. When there is more than one physician code for each death
and the physician identity is known, we can first de-bias the multiple
codes provided from different physicians using the process described in
\citet{insilico}. For the purpose of implementation, we only need to
specify which columns are physician IDs, and which are their coded
causes, respectively.

\begin{Schunk}
\begin{Sinput}
data(SampleCategory)
data(RandomPhysician)
head(RandomPhysician[, 245:250])
\end{Sinput}
\begin{Soutput}
#>   smobph scosts        code1 rev1        code2 rev2
#> 1      .      .          NCD doc9          NCD doc6
#> 2      .      .          NCD doc4          NCD doc3
#> 3      .      .          NCD doc1          NCD doc5
#> 4      .      .      TB/AIDS doc4      TB/AIDS doc7
#> 5      .      .      TB/AIDS doc5      TB/AIDS doc9
#> 6      .      . Communicable doc9 Communicable <NA>
\end{Soutput}
\end{Schunk}

\begin{Schunk}
\begin{Sinput}
doctors <- paste0("doc", c(1:15))
causelist <- c("Communicable", "TB/AIDS", "Maternal",
              "NCD", "External", "Unknown")
phydebias <- physician_debias(RandomPhysician, 
          phy.id = c("rev1", "rev2"), phy.code = c("code1", "code2"), 
          phylist = doctors, causelist = causelist, tol = 0.0001, max.itr = 100)
\end{Sinput}
\end{Schunk}

The de-biased step essentially creates a prior probability distribution
for each death over the broader categories of causes. Then to run
InSilicoVA with the de-biased physician coding, we can simply pass the
fitted object from the previous step to the model. Additional arguments
are needed to specify both the external cause category, since external
causes are handled by separate heuristics, and the unknown category,
which is equivalent to a uniform probability distribution over all other
categories, i.e., the same as the case where no physician coding exists.

\begin{Schunk}
\begin{Sinput}
fit_ins_phy <- codeVA(RandomVA1, model = "InSilicoVA",
             phy.debias = phydebias, phy.cat = SampleCategory, 
             phy.external = "External", phy.unknown = "Unknown",
             Nsim = 10000, auto.length = FALSE) 
\end{Sinput}
\end{Schunk}

Figure \ref{fig:phy3} compares the previous results without including
physicians codes.

\begin{Schunk}
\begin{Sinput}
plotVA(fit_ins_who, title = "Without physician coding")
plotVA(fit_ins_phy, title = "With physician coding")
\end{Sinput}
\begin{figure}[!h]

{\centering \includegraphics[width=1\linewidth,]{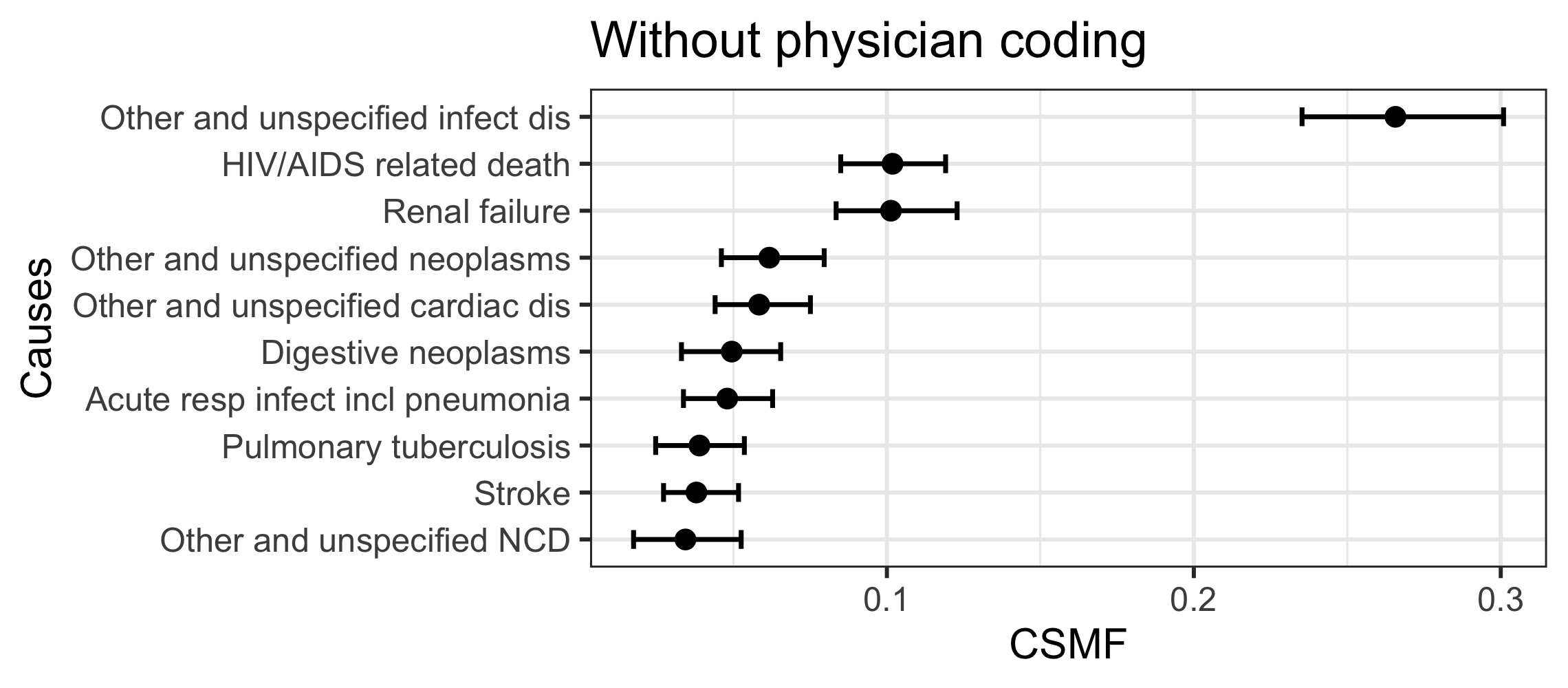} \includegraphics[width=1\linewidth,]{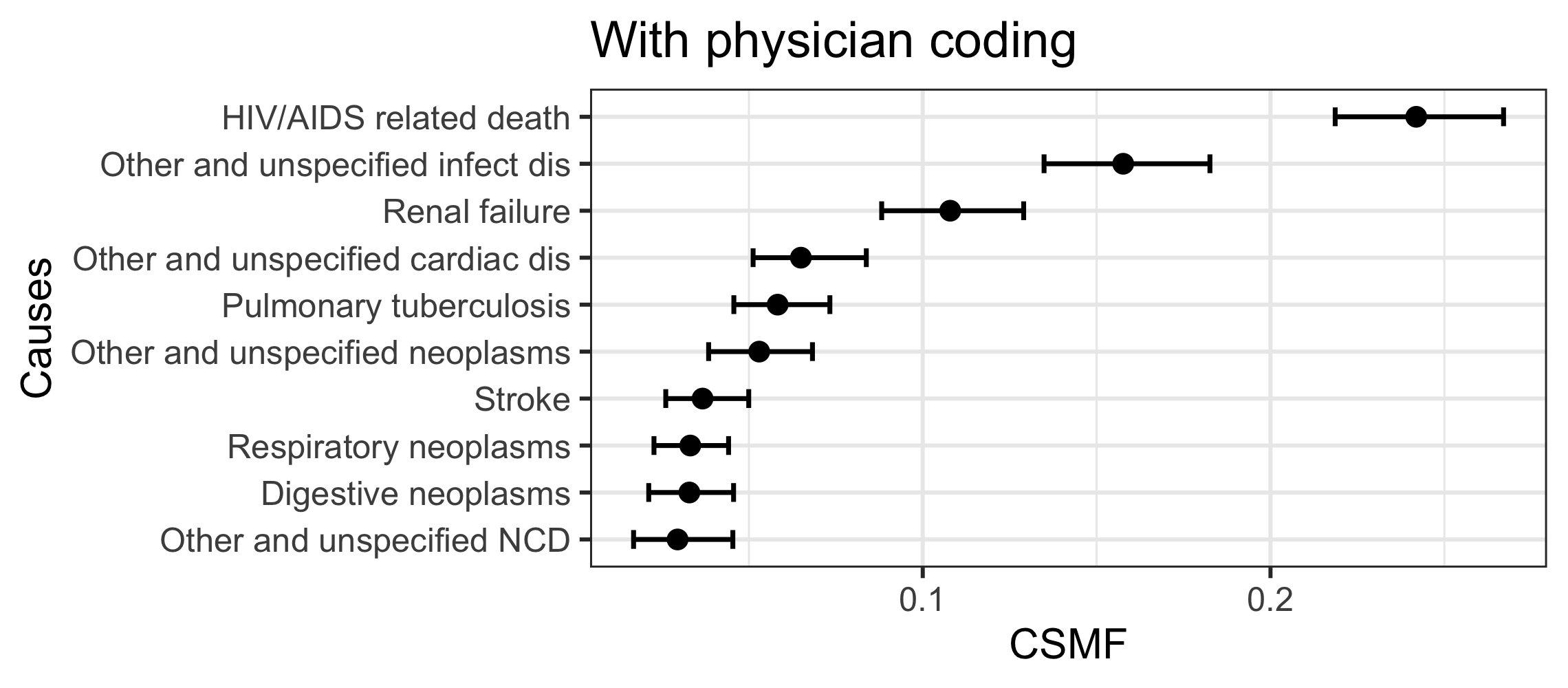} 

}

\caption[Comparing fitted CSMF with and without physicians]{Comparing fitted CSMF with and without physicians}\label{fig:phy3}
\end{figure}
\end{Schunk}

\hypertarget{removal-of-physically-impossible-causes}{%
\subsection{Removal of physically impossible
causes}\label{removal-of-physically-impossible-causes}}

The originally proposed InSilicoVA assumes all causes of death are
possible for each observation. The impact from such an assumption is
mild when data are abundant, but could be problematic when either the
sample size is small or the proportion of missing data is high. In both
cases, physically impossible causes might get assigned with
non-ignorable posterior mass. Since version 1.1.5 of
\CRANpkg{InSilicoVA}, when the input is in the WHO format, the algorithm
automatically checks and removes impossible causes before fitting the
model. The \(k\)-th cause is defined as physically impossible for the
\(i\)-th death if \(P(s_{ij}=1 | y_{i}=k) = 0\) where \(s_{ij}\) is an
indicator that the decedent belongs to a particular sex or age group. We
then consider a cause to be physically impossible for the underlying
population if it is impossible for all the observations in the input
data. For example, with the new implementation, CSMF for
pregnancy-related causes will not be estimated if the input data consist
of only male deaths.

\hypertarget{other-related-software-packages}{%
\section{Other related software
packages}\label{other-related-software-packages}}

\label{sec:other} Since the release of the \CRANpkg{openVA} package on
CRAN, there have been several new developments in both methodology and
software that build on the \CRANpkg{openVA} suite of packages and
further extend its functionalities. Here we briefly survey some of the
related methods, software packages, and ongoing work in the VA community
that are related to \CRANpkg{openVA}.

First, the ability to easily fit and compare existing methods using the
\CRANpkg{openVA} package facilitated the development of several new VA
methods in the last several years. Most of the development focuses on
building statistical models to relax the conditional independence
assumption of symptoms given a cause of death
\citep[e.g.,][]{li2017mix, tsuyoshi2017, moran2021bayesian}. These
methods tend to be computationally more demanding compared to the
algorithms currently included in the \CRANpkg{openVA} package, but
usually provide improved inference. It is future work to include some of
these latest developments in the \CRANpkg{openVA} package for routine
use. Most of these methods have publicly available implementations, such
as the \pkg{farva} package \citep{farva} on GitHub. In another direction
of research, a class of transfer learning methods focuses on building
models to correct for bias in existing methods when predicting out of
domain. These methods take the predicted cause of death assignments and
distributions obtained with the \CRANpkg{openVA} package and learn an
ensemble of the predictions calibrated to the target population
\citep{caliva, fiksel2020generalized}. The \pkg{calibratedVA} package
\citep{calibratedVA} is available to implement these models.

Outside of research community that develops new VA algorithms,
\CRANpkg{openVA} has also been used extensively by governments and
health care organizations, particularly in locations that lack a strong
vital registration system and use VA data to identify the leading causes
of death. To facilitate the work of these users, \CRANpkg{openVA} has
been wrapped into a web application, the openVA App \citep{openva-app},
using the \CRANpkg{shiny} package \citep{shinypkg}. The open source
openVA App is available on GitHub and provides an intuitive interface to
\CRANpkg{openVA} that does not require one to learn R, but provides
visual and tabular output produced by the different VA algorithms. It
also runs the official Tariff algorithm by calling the Python source
code of SmartVA-Analyze \citep{smartVA-git} and processing the output to
be consistent with the other algorithms. The \CRANpkg{openVA} R package
has also been a key component in a larger data analysis pipeline that
pulls VA data from an Open Data Kit (ODK) Collect server and deposits
the assigned causes of death to another server running the District
Health Information Software 2 (DHIS2), which is a common information
management system used in low and middle-income countries. This
open-source software is implemented as a Python package,
openva-pipeline, and is available on GitHub and the Python Package Index
\citep{openva-pipe}. Finally, the R package \CRANpkg{CrossVA}
\citep{crossvapkg} and the Python package pyCrossVA \citep{pyCrossVA}
provide additional toolkits to convert raw VA data from its original
format from the ODK server to the standardized formats discussed before.
Both packages are open source and available on GitHub. The pyCrossVA
package is also available on the Python Package Index.

\hypertarget{conclusion}{%
\section{Conclusion}\label{conclusion}}

In this paper, we introduce the \CRANpkg{openVA} package. This is the
first open-source software that implements and compares the major VA
methods. The \CRANpkg{openVA} package allows researchers to easily
access the most recent tools that were previously difficult to work with
or unavailable. It also enables the compatibility of multiple data input
formats and significantly reduces the tedious work to pre-process
different data formats specific to each algorithm. The software
framework of the \CRANpkg{openVA} package allows for the integration of
new methods developed in the future. The \CRANpkg{openVA} package makes
all the steps involved in analyzing VA data -- i.e., data processing,
model tuning and fitting, summarizing results, and evaluation metrics --
transparent and reproducible. This contributes significantly to the
public health community using VA.

Finally, we make note of several features that will be helpful for
future development. First, many users of the \CRANpkg{openVA} package
may not be familiar with the command line tools or do not have access to
R on their local machines. A well designed graphical user interface can
be very useful in such settings. The work on the shiny web application,
openVA app, is a first step towards making the package more accessible.
The authors intend to extend it to a better front end hosted on secure
centralized servers. Second, although we aim to provide users with all
the available methods for assigning causes of death, there is a lack of
tools for comparing the accuracy and robustness between algorithms.
Thus, much future work is needed to systematically assess, compare, and
combine these methods in a better analytic framework. Finally, the
development of VA algorithms is still an active area of research and it
would be possible to extend the \CRANpkg{openVA} suite to incorporate
better VA algorithms and new types of data such as free-text narratives.

\hypertarget{acknowledgement}{%
\section{Acknowledgement}\label{acknowledgement}}

This work was supported by grants K01HD078452, R01HD086227, and
R21HD095451 from the Eunice Kennedy Shriver National Institute of Child
Health and Human Development (NICHD). Funding was also received from the
Data for Health Initiative, a joint project of Vital Strategies, the CDC
Foundation and Bloomberg Philanthropies, through Vital Strategies. The
views expressed are not necessarily those of the initiative.

\bibliography{openVA-RJ.bib}

\end{article}

\end{document}